\documentclass[11pt]{article}
\usepackage{fullpage,graphicx,hyperref}

\begin{document}

\title{Hybrid Indexes for Repetitive Datasets}
\author{H. Ferrada, T. Gagie, T. Hirvola and S. J. Puglisi}
\date{}
\maketitle

\begin{abstract}
\noindent Advances in DNA sequencing mean databases of thousands of human genomes will soon be commonplace.  In this paper we introduce a simple technique for reducing the size of conventional indexes on such highly repetitive texts.  Given upper bounds on pattern lengths and edit distances, we preprocess the text with LZ77 to obtain a filtered text, for which we store a conventional index.  Later, given a query, we find all matches in the filtered text, then use their positions and the structure of the LZ77 parse to find all matches in the original text.  Our experiments show this also significantly reduces query times.
\end{abstract}

\section{Introduction} \label{sec:introduction}

The British government recently announced plans to sequence the genomes of up to one hundred thousand citizens~\cite{UK12}.  Although sequencing so many people is still a challenge, the cost has dropped sharply in the past few years and the decline is expected to continue.  Now scientists must consider how to store such a massive genomic database in a useful form.

Since human genomes are very similar, this database will be highly repetitive and easy to compress well with, e.g., LZ77~\cite{ZL77}.  Unfortunately, conventional indexes (e.g.,~\cite{LTPS09,LD09,Liu+12}) for approximate pattern matching --- a basic operation in bioinformatics --- are based on FM-indexes~\cite{FM05} or other technologies that do not take good advantage of repetitive structure.  Therefore, these indexes quickly outgrow internal memory when applied to many genomes and must then reside on disk, which slows them down by orders of magnitude.

There are already some experimental LZ- and grammar-based compressed indexes for exact pattern matching (e.g.,~\cite{ANS12,CN12,DJSS12,GGKNP12,KN13,MNSV10,MNKS13}) and these could eventually provide a basis for approximate pattern matching as well (see~\cite{RNOM09}), but science cannot wait.  In this paper we introduce a simple technique, hybrid indexing, for reducing the size of conventional indexes on highly repetitive texts while preserving most or all of their functionality.

Given upper bounds on pattern lengths and edit distances, we preprocess the text with LZ77 to obtain a filtered text, for which we store a conventional index.  Later, given a query, we find all matches in the filtered text, then use their positions and the structure of the LZ77 parse to find all matches in the original text.

We describe our hybrid index in more detail in Section~\ref{sec:main}; this includes details of our implementation\footnote{\href{http://www.cs.helsinki.fi/u/gagie/hybrid}{http://www.cs.helsinki.fi/u/gagie/hybrid}}.  In Section~\ref{sec:experiments} we present experimental results, which show our technique also significantly reduces query times.

\section{Hybrid indexing} \label{sec:main}

Suppose we are given upper bounds on pattern lengths and edit distances, and asked to index a text \(T [1..n]\).   A query to the index consists of a pattern and an edit distance (which is 0 in the case of exact matching).  We store different data structures to be able to find queries' primary matches and secondary matches.  A match for a query is a substring within the given edit distance of the given pattern; it is primary if it contains the first occurrence of a distinct character in $T$ or crosses a phrase boundary in the LZ77 parse of $T$, and secondary otherwise.

To be able to find primary matches, we preprocess the text with LZ77 to obtain a filtered text, which is essentially the subsequence of $T$ containing characters close enough to phrase boundaries in $T$'s LZ77 parse, where ``close enough'' depends on the given upper bounds.  We store a conventional index on this filtered text, and a mapping from it to $T$.  To be able to find secondary matches, we store a data structure by K\"arkk\"ainen and Ukkonen~\cite{KU96}.

Later, given a query, we use the conventional index to find all matches in the filtered text; use the mapping to determine which of those matches correspond to primary matches in $T$; use the mapping again to find those primary matches' positions in $T$; and apply K\"arkk\"ainen and Ukkonen's data structure.

We briefly describe LZ77 in Subsection~\ref{subsec:lz77}.  In Subsection~\ref{subsec:primaries} we give more details about the filtered text and how we find primary matches.  We describe K\"arkk\"ainen and Ukkonen's data structure in Subsection~\ref{subsec:secondaries}.  Finally, in Subsection~\ref{subsec:implementation} we describe details of our implementation.

\subsection{LZ77} \label{subsec:lz77}

We use the variant of LZ77 according to which, for each phrase \(T [i..j]\) in the parse of $T$, either \(i = j\) and \(T [i]\) is the first occurrence of that distinct character, or \(T [i..j]\) occurs in \(T [1..j - 1]\) but \(T [i..j + 1]\) does not occur in \(T [1..j]\).  In the first case, \(T [i]\) is encoded as itself.  In the second case, \(T [i..j]\) is encoded as the pair \((i', j - i + 1)\), where $i'$ is the starting point of the leftmost occurrence of \(T [i..j]\) in $T$; we call this leftmost occurrence \(T [i..j]\)'s source.

For example, if $T$ is
\[\begin{tabular}{l}
\sf 99-bottles-of-beer-on-the-wall-99-bottles-of-beer-\\
take-one-down-and-pass-it-around-98-bottles-of-beer-on-the-wall-\\
98-bottles-of-beer-on-the-wall-98-bottles-of-beer-\\
take-one-down-and-pass-it-around-97-bottles-of-beer-on-the-wall-\\
97-bottles-of-beer-on-the-wall-97-bottles-of-beer-\\
take-one-down-and-pass-it-around-96-bottles-of-beer-on-the-wall- \rm \dots
\end{tabular}\]
then the parse of $T$ (with parentheses around phrases) is
\[\begin{tabular}{l}
({\sf 9}) ({\sf 9}) ({\sf -}) ({\sf b}) ({\sf o}) ({\sf t}) ({\sf t}) ({\sf l}) ({\sf e}) ({\sf s}) ({\sf -}) ({\sf o}) ({\sf f}) ({\sf -b}) ({\sf e}) ({\sf e}) ({\sf r}) ({\sf -o}) ({\sf n}) ({\sf -}) ({\sf t}) ({\sf h}) ({\sf e}) ({\sf -})\\
({\sf w}) ({\sf a}) ({\sf l}) ({\sf l}) ({\sf -}) ({\sf 99-bottles-of-beer-}) ({\sf t}) ({\sf a}) ({\sf k}) ({\sf e-}) ({\sf on}) ({\sf e-}) ({\sf d}) ({\sf o}) ({\sf w}) ({\sf n-}) ({\sf a}) ({\sf n}) ({\sf d})\\
({\sf -}) ({\sf p}) ({\sf a}) ({\sf s}) ({\sf s-}) ({\sf i}) ({\sf t}) ({\sf -a}) ({\sf r}) ({\sf o}) ({\sf u}) ({\sf nd-}) ({\sf 9}) ({\sf 8}) ({\sf -bottles-of-beer-on-the-wall-9})\\
({\sf 8-bottles-of-beer-on-the-wall-98-bottles-of-beer-})\\
({\sf take-one-down-and-pass-it-around-9}) ({\sf 7}) ({\sf -bottles-of-beer-on-the-wall-9})\\
({\sf 7-bottles-of-beer-on-the-wall-97-bottles-of-beer-})\\
({\sf take-one-down-and-pass-it-around-9}) ({\sf 6}) ({\sf -bottles-of-beer-on-the-wall-9}) \dots
\end{tabular}\]
and $T$'s encoding is
\begin{center}
\begin{minipage}[t][15ex]{70ex}
\sf 9 \((1, 1)\) -bot \((6, 1)\) les \((3, 1)\) \((5, 1)\) f \((3, 2)\) \((9, 1)\) \((9, 1)\) r \((11, 2)\) n \linebreak
\((3, 1)\) \((6, 1)\) h \((9, 1)\) \((3, 1)\) wa \((8, 1)\) \((8, 1)\) \((3, 1)\) \((1, 19)\) \((6, 1)\) \((28, 1)\) k \linebreak
\((25, 2)\) \((20, 2)\) \((25, 2)\) d \((5, 1)\) \((27, 1)\) \((21, 2)\) \((28, 1)\) \((21, 1)\) \((60, 1)\) \((3, 1)\) p \linebreak
\((28, 1)\) \((10, 1)\) \((10, 2)\) i \((6, 1)\) \((64, 2)\) \((18, 1)\) \((5, 1)\) u \((66, 3)\) \((1, 1)\) 8 \linebreak
\((3, 30)\) \((85, 49)\) \((51, 34)\) 7 \((3, 30)\) \((199, 49)\) \((51, 34)\) 6 \((3, 30)\) \dots . \hspace{5ex} \linebreak
\end{minipage}
\end{center}

Notice some phrases --- e.g., {\sf 8-bottles-of-beer-on-the-wall-98-bottles-of-beer-} --- overlap their own sources.  Also, while the first verse takes the first four lines of the encoding, the next three verses together take only the last line.  This is typical of LZ77's performance on repetitive inputs.  Finally, although these verses are annoyingly similar, they are much less similar than human genomes.

\subsection{Finding primary matches} \label{subsec:primaries}

Let $M$ and $K$ be the given upper bounds on pattern lengths and edit distances.  Let \(T_{M, K}\) be the text containing the characters of $T$ within distance \(M + K - 1\) of their nearest phrase boundaries; characters not adjacent in $T$ are separated in $T'$ by \(K + 1\) copies of a special character $\#$ not in the normal alphabet.

For example, if $T$ is the example given above then $T_{4, 1}$ is
\[\begin{tabular}{l}
\sf 99-bottles-of-beer-on-the-wall-99-b\,\#\#\,eer-\\
take-one-down-and-pass-it-around-98-bot\,\#\#\,ll-\\
98-bot\,\#\#\,eer-take\,\#\#\,nd-97-bot\,\#\#\,ll-\\
97-bot\,\#\#\,eer-take\,\#\#\,nd-96-bot\,\#\#\,ll- \dots
\end{tabular}\]
or, with parentheses indicating the phrases of $T$,
\[\begin{tabular}{l}
({\sf 9}) ({\sf 9}) ({\sf -}) ({\sf b}) ({\sf o}) ({\sf t}) ({\sf t}) ({\sf l}) ({\sf e}) ({\sf s}) ({\sf -}) ({\sf o}) ({\sf f}) ({\sf -b}) ({\sf e}) ({\sf e}) ({\sf r}) ({\sf -o}) ({\sf n}) ({\sf -}) ({\sf t}) ({\sf h}) ({\sf e}) ({\sf -})\\
({\sf w}) ({\sf a}) ({\sf l}) ({\sf l}) ({\sf -}) ({\sf 99-b\,\#\#\,eer-}) ({\sf t}) ({\sf a}) ({\sf k}) ({\sf e-}) ({\sf on}) ({\sf e-}) ({\sf d}) ({\sf o}) ({\sf w}) ({\sf n-}) ({\sf a}) ({\sf n}) ({\sf d})\\
({\sf -}) ({\sf p}) ({\sf a}) ({\sf s}) ({\sf s-}) ({\sf i}) ({\sf t}) ({\sf -a}) ({\sf r}) ({\sf o}) ({\sf u}) ({\sf nd-}) ({\sf 9}) ({\sf 8}) ({\sf -bot\,\#\#\,ll-9}) ({\sf 8-bo\,\#\#\,eer-})\\
({\sf take\,\#\#\,nd-9}) ({\sf 7}) ({\sf -bot\,\#\#\,ll-9}) ({\sf 7-bo\,\#\#\,eer-}) ({\sf take\,\#\#\,nd-9}) ({\sf 6}) ({\sf -bot\,\#\#\,ll-9}) \dots
\end{tabular}\]

Notice that, for any substring of $T$ with length at most \(M + K\) that contains the first occurrence of a distinct character in $T$ or crosses a phrase boundary in the LZ77 parse of $T$, there is a corresponding and equal substring in $T_{M, K}$.

We do not store $T_{M, K}$ itself explicitly, but we store a conventional index $I_{M, K}$ on $T_{M, K}$.  We assume $I_{M, K}$ can handle queries with pattern lengths up to $M$ and edit distances up to $K$.  Since \(T_{M, K}\) consists of at most \(2 M + 3 K - 1\) characters for each phrase, if $T$ is highly repetitive and $M$ and $K$ are reasonable, then $I_{M, K}$ should be smaller than a conventional index on all of $T$.  Also, for any valid query and any match in $T_{M, K}$, there is at least one match (primary or secondary) in $T$, so querying $I_{M, K}$ should be faster than querying a conventional index for all of $T$.

Let $L$ be the sorted list containing the positions of the first character of each phrase in the parse of $T$, and let $L_{M, K}$ be sorted lists containing the positions of the corresponding characters in $T_{M, K}$.  We store $L$ and $L_{M, K}$.  If \(T [i]\) is the first occurrence of a distinct character in $T$ and \(T_{M, K} [j]\) is the corresponding character in $T_{M, K}$, then we mark $j$ in $L_{M, K}$.

For our example, $L$ is
\[\begin{array}{l}
[1, 2, 3, 4, 5, 6, 7, 8, 9, 10, 11, 12, 13, 14, 16, 17, 18, 19, 21, 22, 23, 24, 25, 26, 27, 28,\\
29, 30, 31, 32, 51, 52, 53, 54, 56, 58, 60, 61, 62, 63, 65, 66, 67, 68, 69, 70, 71, 72, 74,\\
75, 76, 78, 79, 80, 81, 84, 85, 86, 116, 165, 199, 200, 230, 279, 313, 314, \ldots]
\end{array}\]
and $L_{4, 1}$ (with asterisks indicating marked numbers) is
\[\begin{array}{l}
[1^*, 2, 3^*, 4^*, 5^*, 6^*, 7, 8^*, 9^*, 10^*, 11, 12, 13^*, 14, 16, 17, 18^*, 19, 21^*, 22, 23, 24^*,\\
25, 26, 27^*, 28^*, 29, 30, 31, 32, 42, 43, 44^*, 45, 47, 49, 51^*, 52, 53, 54, 56, 57, 58, 59,\\
60^*, 61, 62, 63, 65^*, 66, 67, 69, 70, 71^*, 72, 75, 76^*, 77, 87, 98, 108^*, 109, 119, 130,\\
140^*, 141, \dots]\,.
\end{array}\]

Given the endpoints $i$ and $j$ of a substring \(T_{M, K} [i..j]\) of $T_{M, K}$ that does not include any occurrences of {\sf \#}, we can use $L_{M, K}$ to determine whether the corresponding substring \(T [i'..j']\) of $T$ contains the first occurrence of a distinct character in $T$ or crosses a phrase boundary in the LZ77 parse of $T$.  To do this, we use binary search to find $i$'s successor \(L_{M, K} [s]\).  There are three cases to consider:
\begin{itemize}
\item if \(i < L_{M, K} [s] \leq j\) then \(T [i'..j']\) crosses a phrase boundary;
\item if \(i \leq j < L_{M, K} [s]\) then \(T [i'..j']\) neither contains the first occurrence of a distinct character nor crosses a phrase boundary;
\item if \(i = L_{M, K} [s] \leq j\) then \(T [i'..j']\) contains the first occurrence of a distinct character or crosses a phrase boundary if and only if \(L_{M, K} [s]\) is marked or \(L_{M, K} [s + 1] \leq j\).
\end{itemize}

Also, if \(T [i'..j']\) contains the first occurrence of a distinct character or crosses a phrase boundary, then \(i' = L [s] - L_{M, K} + i\) and \(j' = i' + j - i + 1\).  In other words, we can use $L$ and $L_{M, K}$ as a mapping from $T_{M, K}$ to $T$.

Given a query consisting of a pattern \(P [1..m]\) with \(m \leq M\) and an edit distance \(k \leq K\), we use $I_{M, K}$, $L$ and $L_{M, K}$ to find all primary matches in $T$.  First, we query $I_{M, K}$ to find all matches in $T_{M, K}$.  We then discard any matches containing copies of {\sf \#}.  We use binary search on $L_{M, K}$, as described above, to determine which of the remaining matches correspond to primary matches in $T$.  Finally, we use $L$ and $L_{M, K}$, as described above, to find those primary matches' positions in $T$.

\subsection{Finding secondary matches} \label{subsec:secondaries}

K\"arkk\"ainen and Ukkonen observed that, by definition, any secondary match is completely contained in some phrase.  Also, a phrase contains a secondary match if and only if that phrase's source contains an earlier match (primary or secondary).  It follows that each secondary match is an exact copy of some primary match and, more importantly, once we have found all the primary matches then we can find all the secondary matches from the structure of the LZ77 parse.

To do this, for each primary match \(T [\ell..r]\), we find each phrase \(T [i..j]\) whose source \(T [i'..i' + j - i]\) includes \(T [\ell..r]\) --- i.e., such that \(i' \leq \ell \leq r \leq i' + j - i\).  Notice \(T [\ell'..r'] = T [\ell..r]\), where \(\ell' = i + \ell - i'\) and \(r' = i + 2 \ell - i' - r\).  We record \(T [\ell'..r']\) as a secondary recurrence and recurse on it.

To be able to find quickly all the sources that cover a match, we store a data structure for 2-sided range reporting on the \(n \times n\) grid containing a marker at \((i', j')\) for every phrase's source \(T [i'..j']\).  With each marker we store as a satellite datum the starting point of the actual phrase.  In other words, if a phrase \(T [i..j]\) is encoded as \((i', j - i + 1)\) by LZ77, then there is a marker on the grid at \((i', i' + j - i)\) with satellite datum $i$.

For example, for the phrases shown in Subsection~\ref{subsec:lz77} there are markers at
\begin{center}
\begin{minipage}[t][15ex]{70ex}
(1, 1) (6, 6) (3, 3) (5, 5) (3, 4) (9, 9) (9, 9) (11, 12) (3, 3) (6, 6) (9, 9) \linebreak
(3, 3) (8, 8) (8, 8) (3, 3) (1, 19) (6, 6) (28, 28) (25, 26) (20, 21) (25, 26) \linebreak
(5, 5) (27, 27) (21, 22) (28, 28) (21, 21) (60, 60) (3, 3) (28, 28) (10, 10) \linebreak
(10, 11) (6, 6) (64, 65) (18, 18) (5, 5) (66, 68) (1, 1) (3, 32) (85, 133) \linebreak
(51, 84) (3, 32) (199, 247) (51, 84) (3, 32) \ldots
\end{minipage}
\end{center}
with satellite data
\[\begin{array}{l}
2, 7, 11, 12, 14, 16, 17, 19, 22, 23, 25, 26, 29, 30, 31, 32, 51, 52, 54, 56, 58, 61,\\
62, 63, 65, 66, 67, 68, 70, 71, 72, 75, 76, 78, 79, 81, 84, 86, 116, 165, 200, 230,\\
279, 314, \ldots\,.
\end{array}\]

Notice the markers are simply the encodings of the phrases not consisting of the first occurrences of distinct characters in $T$ (see Subsection~\ref{subsec:lz77}) with each second component \(j - i + 1\) replaced by \(i' + j - i\), where $i'$ is the first component.  Also, the satellite data are the positions of the first characters in those phrases, which are a subset of $L$.

\subsection{Implementation} \label{subsec:implementation}

Recall that, to be able to find primary matches, we store the conventional index $I_{M, K}$ on $T_{M, K}$, the list $L$, and the list $L_{M, K}$.  Because we want our hybrid index to be flexible, we do not consider how $I_{M, K}$ works.  (However, we note that it may sometimes be better to use fewer than \(K + 1\) copies of {\sf \#} as separators, as they serve only to limit the worst-case number of matches in $T_{M, K}$.)

We store $L$ and $L_{M, K}$ using gap coding --- i.e., storing the differences between consecutive values --- with every $g$th value stored un-encoded, where $g$ is a parameter.  We write the differences as \((\lfloor \log_2 d \rfloor + 1)\)-bit integers, where $d$ is the largest difference in the list, and we write the un-encoded values as \((\lfloor \log_2 n \rfloor + 1)\)-bit integers.  To speed up binary search in $L_{M, K}$, we also sample every $b$th value in it, where $b$ is another parameter (typically a multiple of $g$).

Instead of marking values in $L_{M, K}$, we store an array containing the position in $L_{M, K}$ of the first occurrence of each distinct character, in order of appearance.  We note, however, that this array is only necessary if there may be matches of length 1.

To be able to find secondary matches, we store a data structure for 2-sided range reporting on the grid described in Subsection~\ref{subsec:secondaries}.  To build this data structure, we sort the points by their x-coordinates. We store the sorted list $X$ of x-coordinates using gap encoding with every $g$th value stored un-encoded as before, and every $b$th value sampled to speed up binary search.  We store a position-only range-maximum data structure over the list $Y$ of y-coordinates, sorted by x-coordinate.  Finally, we store each satellite datum as a \((\lfloor \log_2 z \rfloor + 1)\)-bit pointer to the cell of $L$ holding that datum, where $z$ is the number of phrases.

We need not store points' y-coordinates explicitly because, if a point has x-coordinate $i'$ and satellite datum $i$, then that point's y-coordinate is \(i' + j - i\), where \(j + 1\) is the value that follows $i$ in $L$.  (We append \(n + 1\) to $L$ to ensure we can always compute $j$ this way, although this is not necessary when the last character of $T$ is a special end-of-file symbol.)  Since we can access $i'$, $i$ and $j$, we can access $Y$.

Once we have found all primary matches, we apply recursive 2-sided range reporting.  To do this, we put the endpoints of the primary matches in a linked list and set a pointer to the head of the list.  Until we reach the end of the list, for each match \(T [\ell..r]\) in the list, we repeat the following procedure:
\begin{enumerate}
\item we use binary search to find $\ell$'s predecessor \(X [k]\) in $X$;
\item we use recursive range-maximum queries to find the values in \(Y [1..k]\) at least $r$;
\item for each point \((i', j')\) we find with \(i' \leq \ell \leq r \leq j'\), we compute $\ell'$ and $r'$ as described in Subsection~\ref{subsec:secondaries};
\item we append the pair \((\ell', r')\) of endpoints to the list and move the pointer forward one element in the list.
\end{enumerate}
When we finish, the list contains the endpoints of all primary matches followed by the endpoints of all secondary matches.

\section{Experiments} \label{sec:experiments}

In our experiments, we compared a hybrid index based on an FM-index for the filtered text, to an FM-index for the original text.  We always used the same implementation of an FM-index\footnote{\href{https://github.com/simongog/sdsl-lite}{https://github.com/simongog/sdsl-lite}} with default parameters.  We used an FM-index instead of a popular index for compressed pattern matching because the latter are usually heavily optimized to take advantage of, e.g., multiple processors; we plan to compare against them after we have parallelized the hybrid index.  We performed our experiments on an Intel Xeon with with 96 GB RAM and 8 processors at 2.4 GHz with 12 MB cache, running Linux 2.6.32-46-server.   We compiled both indexes with {\sf g++} using full optimization.

We used benchmark datasets from the repetitive corpus of the Pizza\&Chili website\footnote{\href{http://pizzachili.dcc.uchile.cl/repcorpus.html}{http://pizzachili.dcc.uchile.cl/repcorpus.html}}.  Specifically, we used the following files:
\begin{description}
\item{\sf cere} --- 37 {\em Saccharomyces cerevisiae} genomes from the Saccharomyces Genome Resequencing Project;\\[-1ex]
\item{\sf einstein} --- versions of the English Wikipedia page on Albert Einstein up to November 10th, 2006;\\[-1ex]
\item{\sf fib41} --- the 41st Fibonacci word $F_{41}$, where \(F_1 = 0\), \(F_1 = 1\), \(F_i = F_{i - 1} F_{i - 2}\);\\[-1ex]
\item{\sf kernel} --- 36 versions of the Linux 1.0.x and 1.1.x kernel.
\end{description}

We generally set \(M = 100\), as that seemed a reasonable value for many applications.  Since standard FM-indexes do not support approximate pattern matching, we set \(K = 0\) throughout.  Based on preliminary tests, we set the sampling parameters $g$ and $b$ for our hybrid index to 32 and 512, respectively.  Notice these parameters have no effect on the FM-indexes.

Table~\ref{tab:sizes} shows the sizes of the uncompressed files, the files compressed with {\sf 7zip}\footnote{\href{http://www.7zip.org}{http://www.7zip.org}} (which does not support pattern matching), the FM-indexes, and the hybrid indexes.  It also shows the construction times for the FM-indexes and hybrid indexes.  The times to build the hybrid indexes include the times to compute the files' LZ77 parses (which, in turn, include the times to build the files' suffix arrays).

\begin{table}[t]
\begin{center}
\begin{tabular}{r|cccc}
& file & {\sf 7zip} & FM & hybrid\\[0.5ex]
\hline\\[-1ex]
{\sf cere} & 440 MB & 5 MB & 88 MB, & 34 MB,\\
& & & 136 s & 1218 s\\[1ex]
{\sf einstein} & 445 MB & 0.3 MB & 26 MB, & 2 MB,\\
& & & 139 s & 161 s\\[1ex]
{\sf fib41} & 256 MB & 0.5 MB & 5 MB, & 0.02 MB,\\
& & & 121 s & 134 s\\[1ex]
{\sf kernel} & 246 MB & 2 MB & 41 MB, & 11 MB,\\
& & & 62 s & 287 s
\end{tabular}
\end{center}
\caption{Sizes of the uncompressed files, the files compressed with {\sf 7zip}, the FM-indexes, and the hybrid indexes; also construction times for the FM-indexes and hybrid indexes.}
\label{tab:sizes}
\end{table}

\begin{figure}[t]
\begin{center}
\includegraphics[width=70ex]{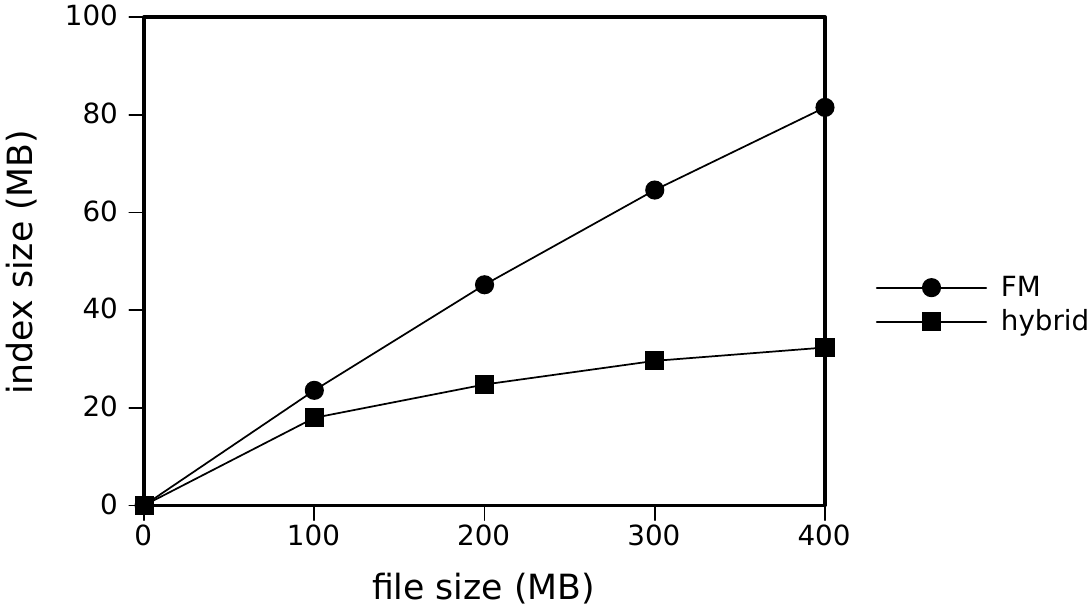}
\end{center}
\caption{Index sizes for prefixes of {\sf cere} of 100, 200, 300 and 400 MB.}
\label{fig:growth}
\end{figure}

To estimate how well hybrid indexing takes advantage of repetitive structure, relative to FM-indexing, we truncated {\sf cere} at 100, 200, 300 and 400 MB, then built FM-indexes and hybrid indexes for those prefixes.  Figure~\ref{fig:growth} shows the sizes of those indexes.

\begin{figure}[t]
\begin{center}
\includegraphics[width=70ex]{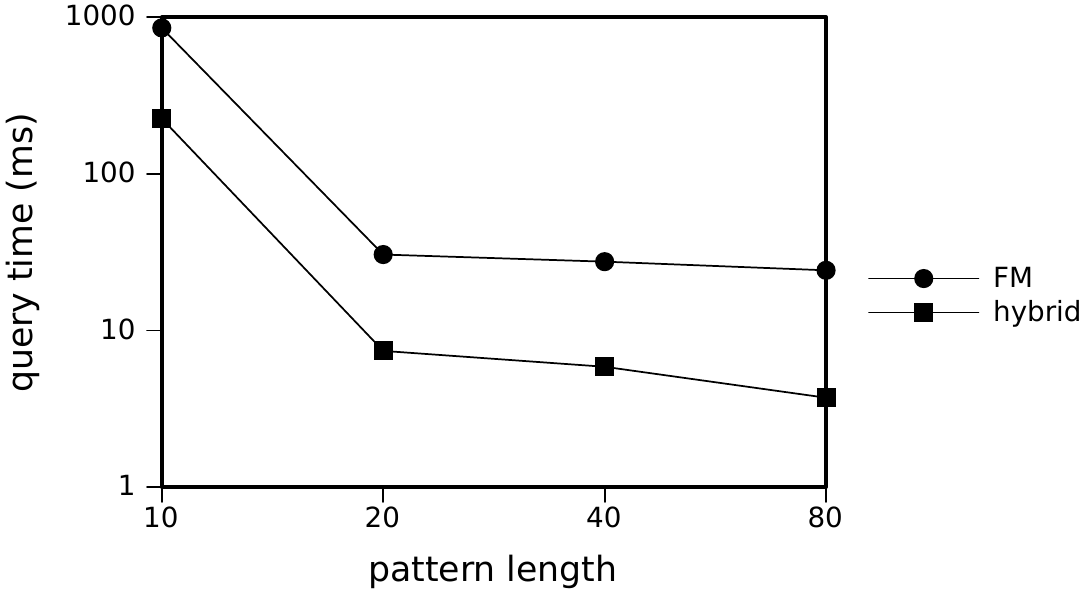}
\end{center}
\caption{Average query times for 3000 patterns.}
\label{fig:queries}
\end{figure}

For 10, 20, 40 and 80, we randomly chose 3000 non-unary substrings that length from {\sf cere} and searched for them with its FM-index and hybrid.  Figure~\ref{fig:queries} shows the average query times, using a logarithmic scale.  The large difference between the query times for patterns of length 10 and those of length 20 seems to be because there are usually far more matches for patterns of length 10; the average time per match stayed roughly the same for all four lengths.

On reflection, it is not surprising that the hybrid index performs well here: while the FM-index finds all matches with its {\sf locate} functionality, the hybrid index finds secondary matches with 2-sided range reporting, which is relatively fast; since {\sf cere} consists of 37 genomes from individuals of the same species, most matches are secondary.

\section{Conclusions and future work} \label{sec:conclusions}

We have introduced a simple technique, hybrid indexing, for reducing the size of conventional indexes on highly repetitive texts.  In our experiments, this technique worked well on benchmark datasets and seemed to scale well.  It also significantly reduced query times.  We plan to optimize our implementation to use, e.g., parallelism across multiple processors; use a more powerful conventional index on the filtered texts; and then compare our hybrid index to popular conventional indexes for approximate pattern matching.

We are also working to optimize hybrid indexing in other ways.  For example, readers may have noticed that, in our example $T_{4, 1}$, there are many copies of {\sf \#\#\,eer-take\,\#\#}.  Including all these copies seems wasteful, since they could be replaced by, e.g., dummy phrases.  We are currently testing whether this noticeably further reduces the size of hybrid indexes in practice.

\section*{Acknowledgments}

Many thanks to Pawe\l\ Gawrychowski, Juha K\"arkk\"ainen, Veli M\"akinen, Gonzalo Navarro, Jouni Sir\'en and Jorma Tarhio, for helpful discussions.


\begin{thebibliography}{10}

\bibitem{UK12}
Strategy for {UK} life sciences: One year on. 2012. See \href{http://www.bis.gov.uk/assets/biscore/innovation/docs/s/12-1346-strategy-for-uk-life-sciences-one-year-on}{http://www.bis.gov.uk/assets/biscore/innovation/docs/s/12-1346-strategy-for-uk-life-sciences-one-year-on}.

\bibitem{ANS12}
D.~Arroyuelo, G.~Navarro, and K.~Sadakane. 2012. Stronger {Lempel-Ziv} based compressed text indexing. {\em Algorithmica}, 62(1--2):54--101.

\bibitem{CN12}
F.~Claude and G.~Navarro. 2012. Improved grammar-based compressed indexes. {\em Proceedings of the 19th Symposium on String Processing and Information Retrieval}, pages 180--192.

\bibitem{DJSS12}
H.~H. Do, J.~Jansson, K.~Sadakane, and W.-K. Sung. 2012. Fast relative {Lempel-Ziv} self-index for similar sequences. {\em Proceedings of the 2nd Conference on Frontiers in Algorithmics and Algorithmic Aspects in Information and Management}, pages 291--302.

\bibitem{Liu+12}
C.-M.~Liu et~al. 2012. {SOAP3}: ultra-fast {GPU}-based parallel alignment tool for short reads. {\em Bioinformatics}, 28(6):878--879.

\bibitem{FM05}
P.~Ferragina and G.~Manzini. 2005. Indexing compressed text. {\em Journal of the ACM}, 52(4):552--581.

\bibitem{GGKNP12}
T.~Gagie, P.~Gawrychowski, J.~K{\"a}rkk{\"a}inen, Y.~Nekrich, and S.~J. Puglisi. 2012. A faster grammar-based self-index. {\em Proceedings of the 6th Conference on Language and Automata Theory and Applications}, pages 240--251.

\bibitem{KU96}
J.~K{\"a}rkk{\"a}inen and E.~Ukkonen. 1996. {Lempel-Ziv} parsing and sublinear-size index structures for string matching. {\em Proceedings of the 3rd South American Workshop on String Processing}, pages 141--155.

\bibitem{KN13}
S.~Kreft and G.~Navarro. 2013. On compressing and indexing repetitive sequences. {\em Theoretical Computer Science}, 483:115--133.

\bibitem{LTPS09}
B.~Langmead, C.~Trapnell, M.~Pop, and S.~L. Salzberg. 2009. Ultrafast and memory-efficient alignment of short {DNA} sequences to the human genome. {\em Genome Biology}, 10(3).

\bibitem{LD09}
H.~Li and R.~Durbin. 2009. Fast and accurate short read alignment with {Burrows-Wheeler Transform}. {\em Bioinformatics}, 25:1754--60.

\bibitem{MNSV10}
V.~M{\"a}kinen, G.~Navarro, J.~Sir{\'e}n and N.~V{\"a}lim{\"a}ki. 2010. Storage and retrieval of highly repetitive sequence collections. {\em Journal of Computational Biology}, 17(3):281--308.

\bibitem{MNKS13}
S.~Maruyama, M.~Nakahara, N.~Kishiue, and H.~Sakamoto. 2013. {ESP}-index: A compressed index based on edit-sensitive parsing. {\em Journal of Discrete Algorithms}, 18:100--112.

\bibitem{RNOM09}
L.~M.~S. Russo, G.~Navarro, A.~L. Oliveira, and P.~Morales. 2009. Approximate string matching with compressed indexes. {\em Algorithms}, 2(3):1105--1136.

\bibitem{ZL77}
J.~Ziv and A.~Lempel. 1977. A universal algorithm for sequential data compression. {\em IEEE Transactions on Information Theory}, 23(3):337--343.

\end{thebibliography}
\end{document}